\documentclass[11pt,a4paper]{article}

\usepackage{jheppub}

\DeclareMathOperator{\Real}{Re}
\DeclareMathOperator{\Imag}{Im}

\title{\boldmath Measurement of $\Gamma_{ee}(J/\psi)$ with KEDR detector}

\author[a]{V.V.~Anashin,}
\author[a,b]{V.M.~Aulchenko,}
\author[a,b]{E.M.~Baldin,} 
\author[a]{A.K.~Barladyan,}
\author[a,b]{A.Yu.~Barnyakov,}
\author[a,b]{M.Yu.~Barnyakov,}
\author[a,b]{S.E.~Baru,}
\author[a]{I.V.~Bedny,}
\author[a,b]{A.E.~Blinov,}
\author[a,b,c]{V.E.~Blinov,}
\author[a,b]{A.V.~Bobrov,}
\author[a,b]{V.S.~Bobrovnikov,}
\author[a,b]{A.V.~Bogomyagkov,}
\author[a,b]{A.E.~Bondar,}
\author[a]{D.V.~Bondarev,}
\author[a,b]{A.R.~Buzykaev,}
\author[a,b]{S.I.~Eidelman,}
\author[a]{Yu.M.~Glukhovchenko,}
\author[a]{V.V.~Gulevich,}
\author[a]{D.V.~Gusev,}
\author[a]{S.E.~Karnaev,}
\author[a]{G.V.~Karpov,}
\author[a]{S.V.~Karpov,}
\author[a,b, 1]{T.A.~Kharlamova, \note{Corresponding author}}
\author[a]{V.A.~Kiselev,}
\author[a,b]{S.A.~Kononov,}
\author[a]{K.Yu.~Kotov,}
\author[a,b]{E.A.~Kravchenko,}
\author[a,b]{V.F.~Kulikov,}
\author[a,c]{G.Ya.~Kurkin,}
\author[a,b]{E.A.~Kuper,}
\author[a,c]{E.B.~Levichev,}
\author[a,b]{D.A.~Maksimov,}
\author[a]{V.M.~Malyshev,}
\author[a,b]{A.L.~Maslennikov,}
\author[a]{A.S.~Medvedko,}
\author[a,b]{O.I.~Meshkov,}
\author[a]{S.I.~Mishnev,}
\author[a,b]{I.I.~Morozov,}
\author[a,b]{N.Yu.~Muchnoi,}
\author[a]{V.V.~Neufeld,}
\author[a]{S.A.~Nikitin,}
\author[a,b]{I.B.~Nikolaev,}
\author[a]{I.N.~Okunev,}
\author[a,b,c]{A.P.~Onuchin,}
\author[a]{S.B.~Oreshkin,}
\author[a]{I.O.~Orlov,}
\author[a,b]{A.A.~Osipov,}
\author[a,b]{S.V.~Peleganchuk,}
\author[a,c]{S.G.~Pivovarov,}
\author[a]{P.A.~Piminov,}
\author[a]{V.V.~Petrov,}
\author[a]{A.O.~Poluektov,}
\author[a]{I.N.~Popkov,}
\author[a,b]{V.G.~Prisekin,}
\author[a,b]{O.L.~Rezanova,}
\author[a,b]{A.A.~Ruban,}
\author[a]{V.K.~Sandyrev,}
\author[a]{G.A.~Savinov,}
\author[a,b]{A.G.~Shamov,}
\author[a]{D.N.~Shatilov,}
\author[a,b]{B.A.~Shwartz,}
\author[a]{E.A.~Simonov,}
\author[a]{S.V.~Sinyatkin,}
\author[a]{Yu.I.~Skovpen,}
\author[a]{A.N.~Skrinsky,}
\author[a]{V.V.~Smaluk,}
\author[a,b]{A.V.~Sokolov,}
\author[a,b]{A.M.~Sukharev,}
\author[a,b]{E.V.~Starostina,}
\author[a,b]{A.A.~Talyshev,}
\author[a,b]{V.A.~Tayursky,}
\author[a,b]{V.I.~Telnov,}
\author[a,b]{Yu.A.~Tikhonov,}
\author[a,b]{K.Yu.~Todyshev,}
\author[a]{G.M.~Tumaikin,}
\author[a]{Yu.V.~Usov,}
\author[a]{A.I.~Vorobiov,}
\author[a]{A.N.~Yushkov,}
\author[a,b]{V.N.~Zhilich,}
\author[a,b]{V.V.~Zhulanov,}
\author[a,b]{A.N.~Zhuravlev}

\affiliation[a]{Budker Institute of Nuclear Physics, 11, akademika
 Lavrentieva prospect,  Novosibirsk, 630090, Russia}
\affiliation[b]{Novosibirsk State University, 2, Pirogova street,  Novosibirsk, 630090, Russia}
\affiliation[c]{Novosibirsk State Technical University, 20, Karl Marx
  prospect,  Novosibirsk, 630092, Russia}

\emailAdd{T.A.Kharlamova@inp.nsk.su}

\date{}

\abstract{
The product of the electronic width of the $J/\psi$ meson and
the branching fractions of its decay to hadrons and
electrons has been measured using the KEDR detector at the VEPP-4M 
 $e^+e^-$ collider. The obtained values are
\begin{gather*} 
    \Gamma_{ee}(J/\psi) = 5.550 \pm 0.056 \pm 0.089 \, \text{keV}, \\
    \Gamma_{ee}(J/\psi) \cdot \mathcal{B}_\text{hadrons}(J/\psi) = 4.884 \pm
    0.048 \pm 0.078 \, \text{keV}, \\
    \Gamma_{ee}(J/\psi) \cdot \mathcal{B}_{ee}(J/\psi) = 0.3331 \pm
    0.0066 \pm 0.0040  \, \text{keV}.
\end{gather*}
The uncertainties shown are statistical and systematic, respectively.
Using the result presented and the world-average value of the electronic
branching fraction, one obtains the total width of the $J/\psi$ meson:
\begin{equation*}
\begin{split}
\Gamma & = 92.94 \pm 1.83 \, \text{keV}.
\end{split}
\end{equation*}
These results are consistent with the previous experiments.
}

\begin{document}
\maketitle
\flushbottom

\section{Introduction}
\label{sec:intro}

The $J/\psi$ resonance, a bound state of $c\bar{c}$ quarks, was discovered
more than forty years ago but its investigation is still
actual. Fundamental properties of this meson including the branching
fractions of leptonic and hadronic decays are  
important for understanding the quarkonium decay dynamics. The leptonic 
width of the $J/\psi$ meson is used in calculations
of  c-quark mass~\cite{mc1,mc2} and the hadronic
contribution to the muon \mbox{$g-2$~\cite{gm2}}. 
It is also used for various
calculations of radiative corrections
due to the vacuum polarization and the initial-state
radiation. The current precision of $\Gamma_{ee}$ in the 
potential models and in the lattice calculations~\cite{laticeQCD, etmccharm}
 is compatible with that of the  
world-average value~\cite{pdg2014} and increase of the experimental 
precision for this value can be crucial for further development 
of the LQCD calculation techniques.

Measurements of the $J/\psi$ widths  
have a long history.   They were studied at MarkI~\cite{mark1} and  
ADONE~\cite{frascati}, and later at BES~\cite{bes95}, BaBar~\cite{babar04}, 
CLEO~\cite{cleo06}, KEDR~\cite{KEDR:2010, KEDR:2014} and \mbox{BESIII}~\cite{bes16}. Usually
$\Gamma_{ee}$  is measured in $J/\psi$ decays to hadrons, $e^+e^-$ or $\mu^+\mu^-$
final states and the obtained value is the product of $\Gamma_{ee}$ 
to the corresponding  branching fraction.
At present the best accuracy in the
determination of $\Gamma_{ee}$ has been obtained by the \mbox{BESIII} 
collaboration~\cite{bes16} based on
the $\Gamma_{ee}\cdot\mathcal{B}_{\mu\mu}(J/\psi)$ measurement in the 
initial-state radiation process $e^+e^- \to J/\psi \gamma \to
\mu^+\mu^-\gamma$ and ${B}_{\mu\mu}(J/\psi)$ branching
fraction~\cite{pdg2014}. The best accuracy of the $\Gamma_{ee} \cdot \mathcal{B}_{\text{hadrons}}$
value has been reached by combining the result on 
$\Gamma_{ee}$~\cite{pdg2014} with $\mathcal{B}_{\text{hadrons}}$ from BES~\cite{bes95}.

This work continues a series of experiments on measuring properties of
charmonium resonances performed by the KEDR collaboration~\cite{KEDR:2010,KEDR:2014,KEDR:2003,KEDR:2012,KEDR:2015}. 
In 2010 partial widths $\Gamma_{ee}\cdot\mathcal{B}_{ee}(J/\psi)$ and
$\Gamma_{ee}\cdot\mathcal{B}_{\mu\mu}(J/\psi)$ were measured with high
accuracy of 2.4\% and 2.5\%,  respectively~\cite{KEDR:2010}. 
In this
article we present new results on $\Gamma_{ee}$ and $\Gamma_{ee} \cdot
\mathcal{B}_{\text{hadrons}}$ obtained by measuring the cross sections
of $e^+e^- \to \text{hadrons}$ and  $e^+e^-  \to e^+e^- $
as a function of the centre-of-mass (c.m.) energy in the vicinity of the $J/\psi$ 
resonance with the
KEDR detector at the VEPP-4M $e^+e^-$ collider.

\section{Experiment and data sample}
\label{sec:experiment}

\begin{figure*}[t!] 
\centering\includegraphics*[width=0.9\textwidth]{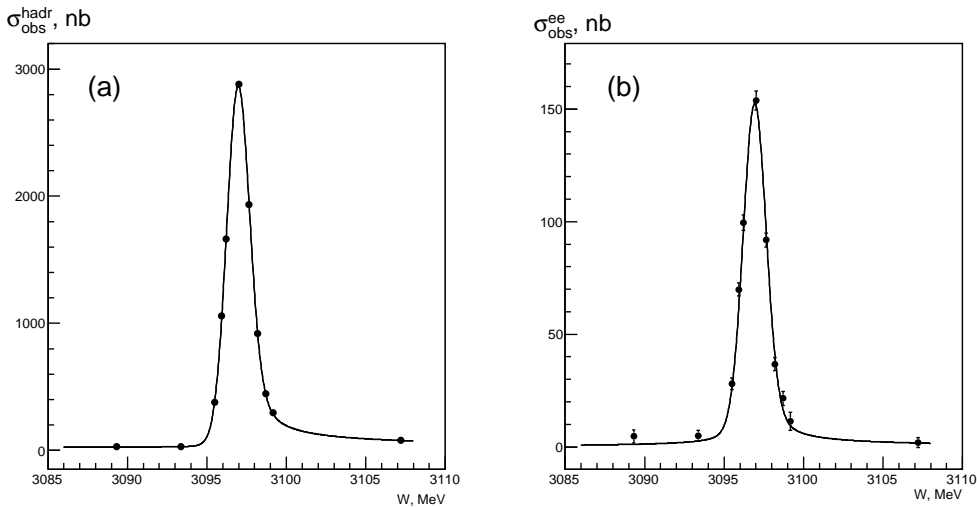}
\caption{Observed cross section as a function
of the c.m. energy (a) for $e^+e^- \to J/\psi \to$ hadrons and 
(b) for $e^+e^- \to J/\psi \to e^+ e^-$ processes. The
curves are the result of the combined fit,\  $\chi^2$ of the fit equals
6.6 and 9.0 for the hadronic and leptonic channels, respectively, with the
total number of degrees of freedom 15.
}
\label{cross}
\end{figure*}
A detailed description of the VEPP-4M $e^+e^- $ collider 
and the  KEDR detector can be found in Refs.~\cite{vepp,kedr}.
Our analysis is based on the same data set, with an 
integrated luminosity of $230~\text{nb}^{-1}$,  
as that used in the KEDR analysis of the leptonic channels~\cite{KEDR:2010}.
The data sample corresponds to 250 thousands of produced $J/\psi$ mesons. 
During the scan the data were collected at 11 energy points as shown 
in Fig.~\ref{cross} that allows a fit of the resonance shape and 
determination of the nonresonant background contributions to be performed. 
The beam energy was measured by the resonant depolarization
method~\cite{rdm:Bukin}.  26
calibrations were carried out during the scan, before and after data
taking at each energy point. Between the calibrations the beam energy was
interpolated with the accuracy better than 15 keV.

\section{$e^+e^-$  cross section in vicinity of a narrow resonance}
\label{sec:crosssection}

The cross section for the annihilation process $e^+e^- \to \text{hadrons}$
in the vicinity of a narrow resonance can be presented in the 
form~\cite{KEDR:2012}:
\begin{equation}
\begin{split} 
&\sigma^{\rm{hadr}}_{\rm{n.r.}}(W) = \frac{12\pi} { W^2 }
 \Bigg\{ \bigg(1+\delta_{\mathrm{sf}}\bigg)
\Bigg[
  \frac{\Gamma_{ee}\tilde{\Gamma}_{\rm h}}{\Gamma M} 
\Imag{f(W)} 
 -\,\frac{2 \alpha\sqrt{R\,\Gamma_{ee}\tilde{\Gamma}_{\rm h}\,}}{3 W}\,
\lambda\,\Real{\frac{f^{*}(W)}{1\!-\!\Pi_0}}\
\Bigg] \\
&-\,\frac{\beta\,\Gamma_{ee}\tilde{\Gamma}_{\rm h}}{2 \Gamma M}\,
\Bigg[
 \Bigg(1\!+\!\frac{M^2}{W^2}\Bigg)\,
\arctan{\frac{\Gamma W^2}{M(M^2\!-\!W^2\!+\!\Gamma^2)}}
-\frac{\Gamma M}{2 W^2}
  \ln{\frac{\bigg(\frac{M^2}{W^2}\bigg)^2+\bigg(\frac{\Gamma
        M}{W^2}\bigg)^2}{\bigg(1-\frac{M^2}{W^2}\bigg)^2+\bigg(\frac{\Gamma
        M}{W^2}\bigg)^2}} \Bigg]\,
 \Bigg\}\,,
\end{split} 
\label{BWrelativistic}
\end{equation}
where $W$ is the c.m. energy, $M$ is the mass of the resonance,  
$\Gamma$ is its
total width, $\alpha$ is the fine structure constant and 
$R$ is the ratio $\sigma(e^+e^-  \to {\rm hadrons})/\sigma(e^+e^-  \to \mu^{+}\mu^{-})$ 
outside of the resonance region. The truncated vacuum-polarization 
operator $\Pi_0$ 
does not include a contribution of the resonance itself. 

The radiative correction $\delta_{\text{sf}}$ can be obtained from the
structure-function approach of Ref.~\cite{KF}:
\begin{equation}\label{eq:deltasf}
  \delta_{\rm sf}=\frac{3}{4}\beta+
   \frac{\alpha}{\pi}\left(\frac{\pi^2}{3}-\frac{1}{2}\right)+
  \beta^2\left(\frac{37}{96}-\frac{\pi^2}{12}-
  \frac{1}{36}\ln\frac{W}{m_{\rm e}} \right)\,,
\end{equation}
\begin{equation}
\beta = \frac{4\alpha}{\pi} \left( \ln\frac{W}{m_{\rm e}} -\frac{1}{2}\right)\,,
\label{eq:beta}
\end{equation}
where $m_{\rm e}$ is the electron mass. The function $f$ is defined as
\begin{equation}
f(W) =  \frac{\pi\beta}{\sin{\pi\beta}}\,
   \Bigg(\frac{W^2}{M^2-W^2-i M \Gamma} \Bigg)^{1-\beta}
\!\!\!\!\!.
\label{resfunction}
\end{equation}

The parameter $\lambda$ in Eq.~\eqref{BWrelativistic} characterizes 
the strength of 
the interference effect in the inclusive hadronic cross section. According to Ref.~\cite{KEDR:2012} 
the expression for $\lambda$ can be written as
\begin{equation}
  \lambda = \sqrt{\frac{R \mathcal{B}_{ee}}{\mathcal{B}_{\rm hadrons}}} + \sqrt{\frac{1}{\mathcal{B}_{\rm hadrons}}}\,
        \sum\limits_m\!
    \sqrt{b_m \mathcal{B}^{(s)}_m\,}
          \left<\cos{\phi_m}\right>_{\Theta}\,.
\label{eq:lambdaSum}
\end{equation}
The summation is performed over all exclusive hadronic modes.

Here and below $\left<\cos{\phi_m}\right>_{\Theta}$ and
$\left<\sin{\phi_m}\right>_{\Theta}$  are the cosine and sine of the relative 
phase of the strong and electromagnetic amplitudes for the mode $m$
averaged over the phase space of the products, 
$b_m\!=\!R_m/R$ is the branching fraction of the corresponding continuum
process, $\mathcal{B}_{ee}$ is a probability of  the decay to an $e^+e^-$ pair,
$\mathcal{B}_{\rm hadrons}$ is the total decay probability to hadrons
and $\mathcal{B}^{(s)}_m\!=\!\Gamma^{(s)}_m/\Gamma$, where $\Gamma^{(s)}$
is the contribution of the strong interaction to the partial width for
the mode $m$.

Due to the resonance -- continuum interference the effective hadronic
width $\tilde{\Gamma}_{\rm h}$ can differ from the true
hadronic partial width $\Gamma_{\rm hadrons} = \sum\limits_{m} \Gamma_{m}$:

\begin{equation}
\begin{split}  
&\tilde{\Gamma}_{\rm h} =\Gamma_{\rm hadrons} \,\,\times  \left(1+\frac{2\alpha}{3(1-\Real
   \Pi_0) \mathcal{B}_{\rm hadrons}} \sqrt{\frac{R}{\mathcal{B}_{ee}}}\,
   \! \times 
 \sum\limits_m\!\sqrt{b_m \mathcal{B}^{(s)}_m}
       \left<\sin{\phi_m}\right>_{\Theta}\right)\,. 
\label{eq:Gh}
\end{split}  
\end{equation}

In this analysis it was assumed that the relative phases of the strong and
electromagnetic amplitudes in different decay modes are not correlated.
Consequences and experimental verification of this assumption  
are discussed in detail in Refs.\cite{KEDR:2012,KEDR:2015}.

The differential $e^+e^- $ cross section is calculated with
\begin{equation}
  \begin{split}  
    &\left(\frac{d\sigma}{d\Omega}\right)^{ee\to ee} \!\!\! =
    \left(\frac{d\sigma}{d\Omega}\right)_{\text{QED}}^{ee\to ee}+
    \!\!\!\!\!\!\quad\frac{1}{W^2}\left(1\!+\!\delta_{sf}\right)
    \left\{\,\frac{9}{4}\frac{\Gamma^2_{ee}}{\Gamma
      M}(1+\cos^2\theta) \Imag f -\right.\\
    &  \!\!\!\!\!\!\!\!\!\!\qquad\left.\frac{3\alpha}{2}\frac{\Gamma_{ee}}{M}
    \left [(1+\cos^2\theta)\Real \frac{f^{*}}{1\!-\!\Pi_0(s)}-
    \frac{(1+\cos\theta)^2}{(1-\cos\theta)}
              \Real \frac{f^{*}}{1\!-\!\Pi_0(t)}\right ]
    \right\}\,,     
\label{eetoee}
\end{split}
\end{equation}
where  $s=W^2$ and
$t=- W^2 \cdot (1 - cos\,\theta)/2$ are the c.m. energy
squared and momentum transfer squared, $\theta$ is the electron scattering angle.
The first term in Eq.~\eqref{eetoee} represents the QED cross section obtained with
the Monte Carlo technique~\cite{bhwide,mcgpj}.
The second term is responsible for the resonance contribution and the third
one for the interference.
The accuracy of the
formulae~\eqref{eetoee} about 0.1\% is sufficient for this work and is
confirmed with more precise expressions given in~\cite{chee}. 

\section{Data analysis}
\label{sec:analysis}

\subsection{MC simulation}
\label{subsec:MC}

We used MC samples of $J/\psi$ inclusive decays and the continuum
multihadron events to obtain the detector efficiency. 
 The samples were generated with the tuned version of the BES generator~\cite{besgen} based 
on JETSET 7.4~\cite{jetset}. The procedure
of the parameter tuning is discussed in detail in
Sec.~\ref{subsec:simulation}.
The generated events were
reweighted to ensure that the branching fractions of the most probable
decay modes correspond to the results of the PDG fit~\cite{pdg2014}.
 MC samples of Bhabha events required for the luminosity determination
were simulated using the \mbox{BHWIDE}~\cite{bhwide} and \mbox{MCGPJ}~\cite{mcgpj} generators. 
Generated MC events were then processed with the detector simulation
package based on GEANT, version 3.21~\cite{geant}, and reconstructed
with the same conditions as experimental data.

During the data taking in 2005 there was an additional online condition
-- the number of hits in the vertex detector (VD) should not exceed 60 which
corresponded to 10 charged tracks.
Due to substantial crosstalk in VD electronics, there was some loss of signal
events. The effect of crosstalk was carefully simulated.

To take into account the signal and background coincidences, a trigger
from arbitrary beam crossings was implemented. The events recorded with
this "random trigger" were superimposed with simulated events.

\subsection{Trigger requirements}
\label{subsec:trigger}

The trigger consists of two hardware levels:
the primary trigger (PT) and the secondary trigger (ST)~\cite{TALYSHEV}.
The primary trigger required signals from two or more non-adjacent
scintillation counters  or an energy deposition
in the endcap calorimeter of at least 100~MeV. A veto from  CsI
calorimeter crystals closest to
the beam line was used to
suppress the machine background.
The conditions of the secondary trigger were rather complicated, and 
were satisfied by events with two tracks in the vertex detector and 
the drift chamber
or with a single track which deposited more the 70 MeV in the barrel
calorimeter.

During the offline analysis all events (both recorded in experiment and 
simulated) were
required to pass through the software event filter. It used a
digitized response from detector subsystems and applied tighter conditions
on its input in order to decrease the effect of calorimeter energy threshold
and possible hardware-trigger instability.

\subsection{Luminosity determination}
\label{subsec:lum}

For the absolute luminosity determination, $e^+e^-$ events in the barrel LKr
calorimeter~\cite{kedr} were used taking into account the contribution of  $J/\psi$  
decays into $e^+e^-$ (see Eq.~\eqref{eetoee}).
 
The final-state radiation (FSR) effects are considered using the 
PHOTOS package~\cite{photos}. 
The $J/\psi \to e^+e^-$ cross section is shown in Fig.~\ref{cross}b obtained 
by subtracting
the contribution of  Bhabha events from the total $e^+e^- \to e^+e^-$ cross section.

The $e^+e^- $ event selection includes the following criteria in addition to
trigger requirements:
\begin{itemize}
\item two clusters within the polar angle range   $40 < \theta
  <140^{\circ}$ and the energy $E_{\rm 1,2}$ larger than 700 MeV each;
\item the energy deposition outside of  those two clusters 
smaller  than  $10\%$ of the total energy deposited in the calorimeter
$E_{\rm cal}$;
\item acollinearities of the polar $\Delta\theta$ and azimuthal $\Delta\varphi$ angles smaller than~15$^{\circ}$;
\item event sphericity $S_{\rm ch}$ 
calculated with
  charged particles smaller than 0.05;
\item two or three tracks in the drift chamber coming from the interaction 
point: 
the impact parameter with respect to the beam axis $\rho < 0.5 \ \text{cm}$, 
the coordinate of the
point of closest approach $|\text{z}_0| < 13 \ \text{cm}$ 
and the transverse momentum
  $P_{\rm t} > 100~\text{MeV}$.
\end{itemize}
Cosmic background was additionally suppressed with the muon
system by veto signals from opposite or 
adjacent to opposite octants or
more than three layers fired in one octant.  Alternatively, cosmic events
were suppressed with the time-of-flight condition.

\begin{figure*}[htb!]
\begin{center}
\includegraphics*[width=0.49\textwidth]{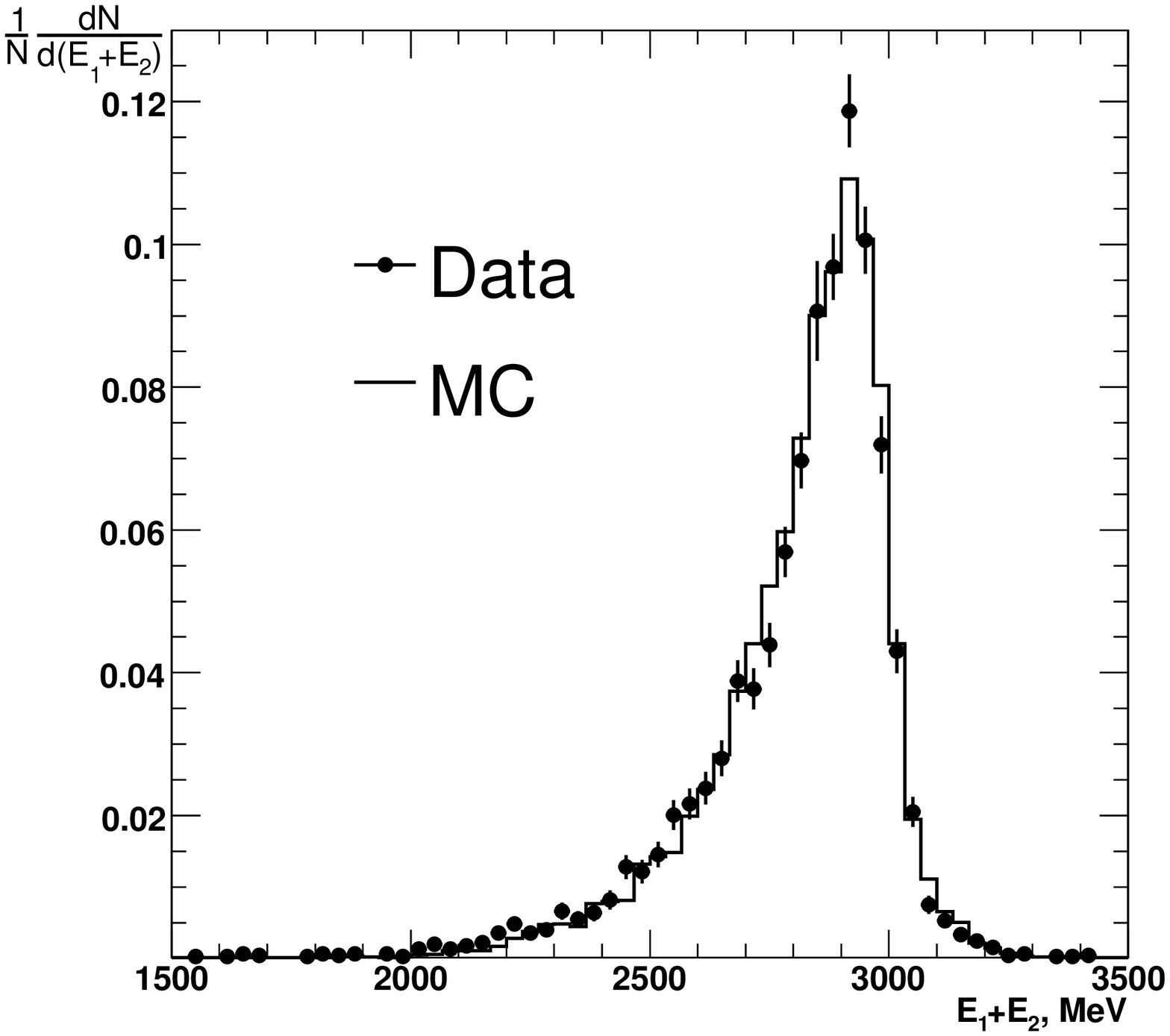}
\hfill
\includegraphics*[width=0.49\textwidth]{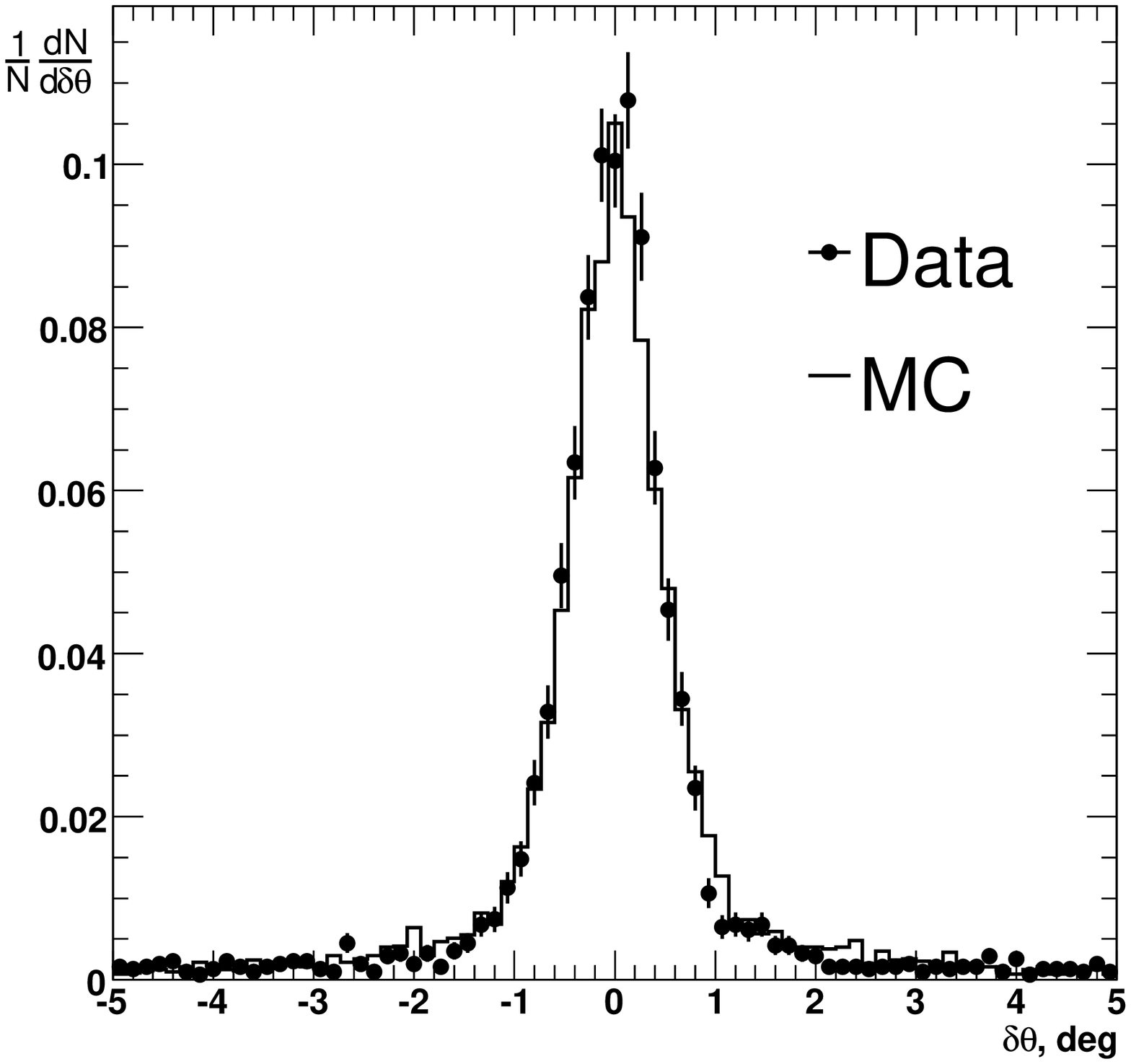}
\caption{Properties of $e^+e^-$ events produced at the $J/\psi$ peak - sum 
of two maximum cluster energies and polar-angle acollinearity in degrees. 
All distributions are normalized to unity.}
\label{eesel1}
\end{center}
\end{figure*}

\begin{figure*}[htb!]
\begin{center}
\includegraphics*[width=0.8\textwidth]{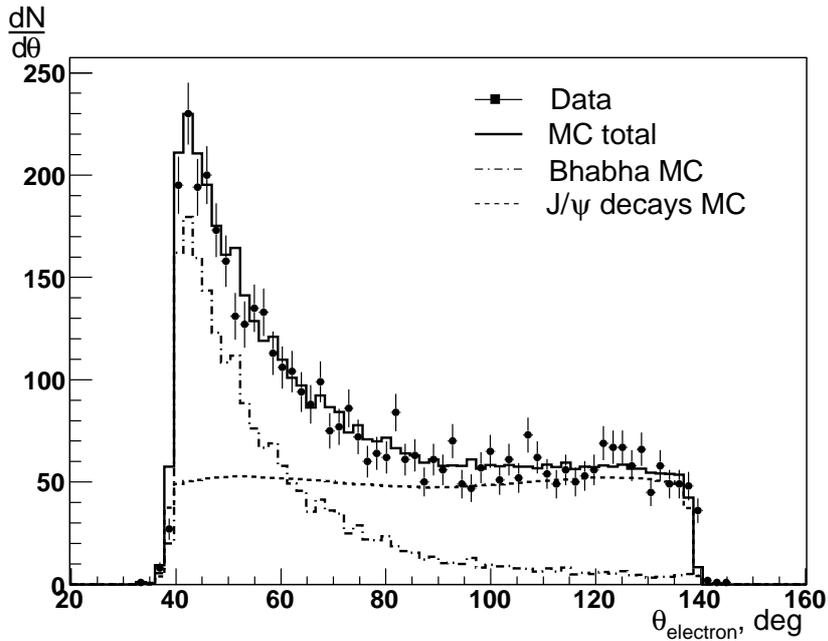}
\caption{Distribution of the electron polar angle at the $J/\psi$ peak. 
The points show experimental data. The histograms
  correspond to MC simulation: the dashed-doted histogram represents Bhabha scattering, the dashed histogram represents a contribution
of the $J/\psi$ resonance and their interference calculated according to Eq.~\eqref{eetoee},  and the solid-line histogram
  is the sum of the  contributions.
}
\label{theta}
\end{center}
\end{figure*}

Figure \ref{eesel1} shows comparison between $e^+e^- \to e^+e^-$
data and MC simulation. The distribution in the electron scattering angle
for selected $e^+e^-$ events is shown in
Fig.~\ref{theta}. The angular distributions of events from Bhabha
scattering and from  $J/\psi$  decay are different which allows us to separate those
contributions at each data point. 

\subsection{Selection of hadronic events}
\label{subsec:hadrsel}

In our analysis the following selection requirements are applied:
\begin{itemize}
\item total energy deposition in the calorimeter $700 < E_{\rm cal} < 2500$
  MeV;
\item more than 15\% of total energy  deposited in the barel LKr
  calorimeter $E_{\rm LKr}/E_{\rm cal}>0.15$;
\item at least one track with $\rho < 0.5 \ \text{cm}$,  $|\text{z}_0| <
  13 \ \text{cm}$ and $P_{\rm t} > 100 \  \text{MeV}$;
\item at least three particles in the detector, including tracks in
  the drift chamber and calorimeter clusters, which are not
  associated with any track;
\item the ratio of the Fox-Wolfram moments~\cite{fox} $H_2/H_0 < 0.9$.
\end{itemize}

The requirements on energy deposition separate hadronic events from
backgrounds: the upper requirement reduces a fraction of $e^+e^-$ events and 
the lower one suppresses  $\mu^+\mu^-$ and machine backgrounds. 
The requirement on the ratio of the Fox-Wolfram moments $H_2/H_0$ is significant
in reducing background from quasi-collinear $e^+e^-$  events with
additional particles from radiation and interaction with detector material. 
Cosmic events were additionally suppressed as in selection of $e^+e^-$ events.

\begin{figure*}[htb!] 
\includegraphics*[width=0.49\textwidth,height=0.25\textheight]{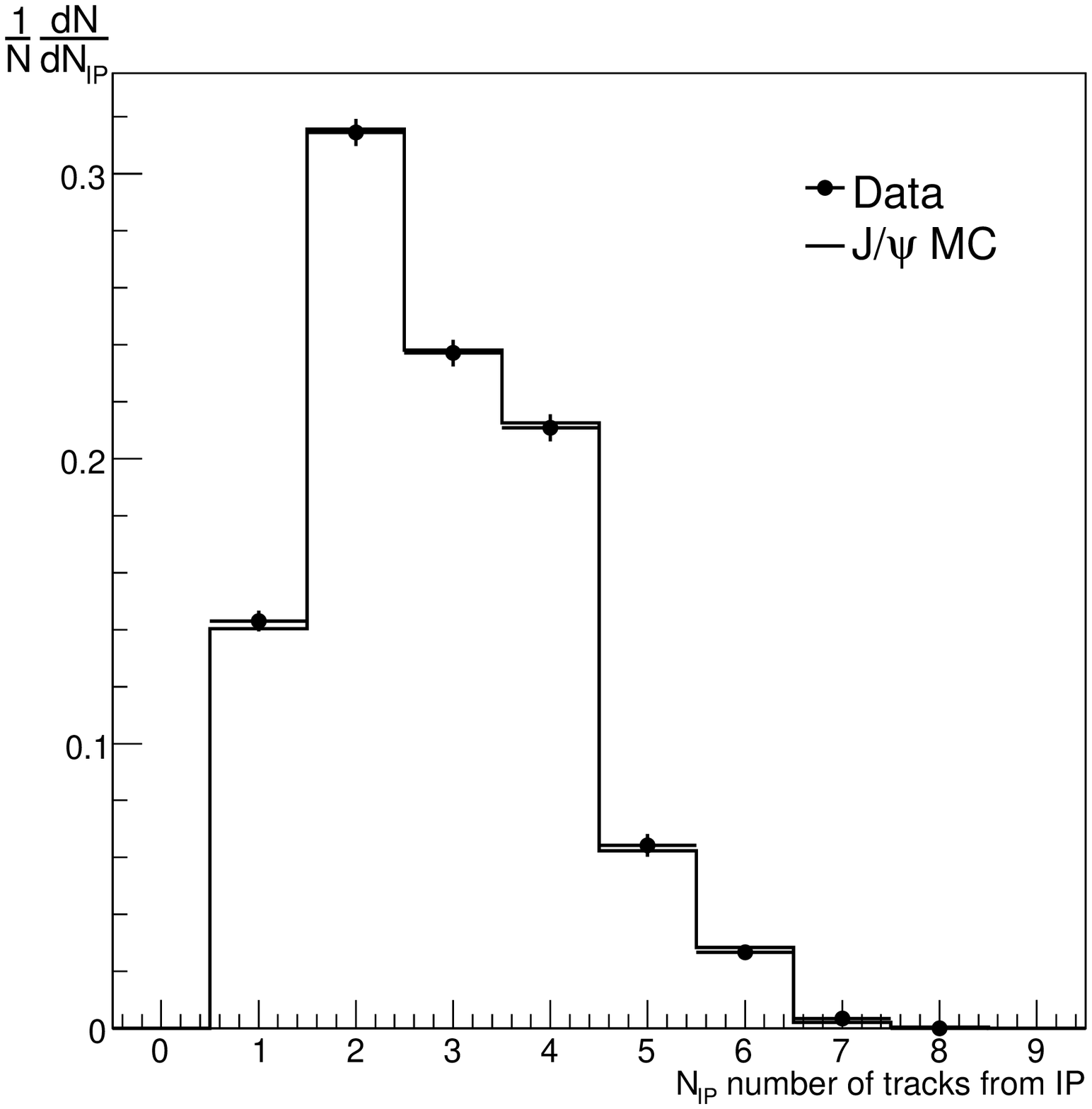}
\hfill
\includegraphics*[width=0.49\textwidth,height=0.25\textheight]{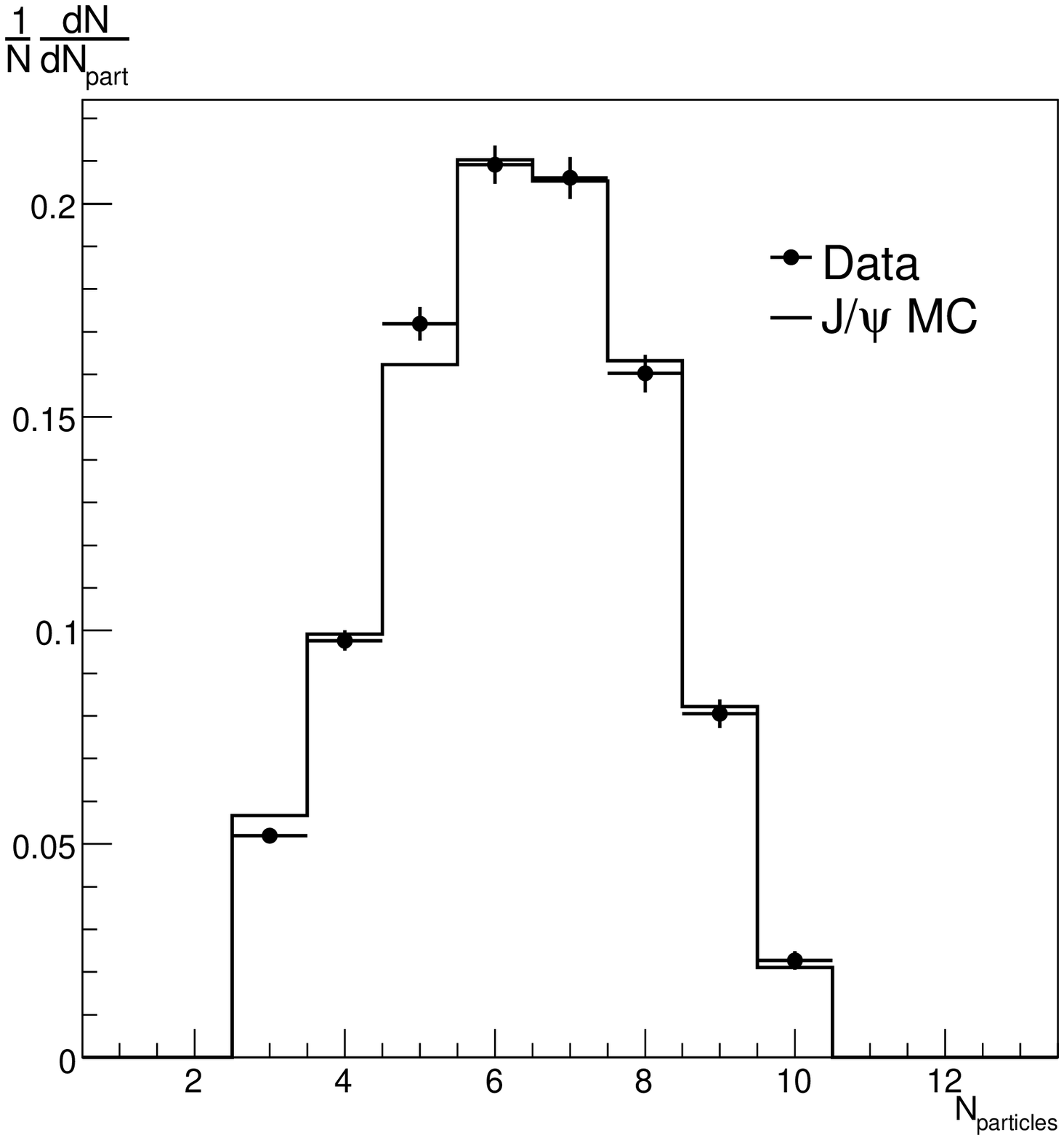}\\
\includegraphics*[width=0.49\textwidth,height=0.25\textheight]{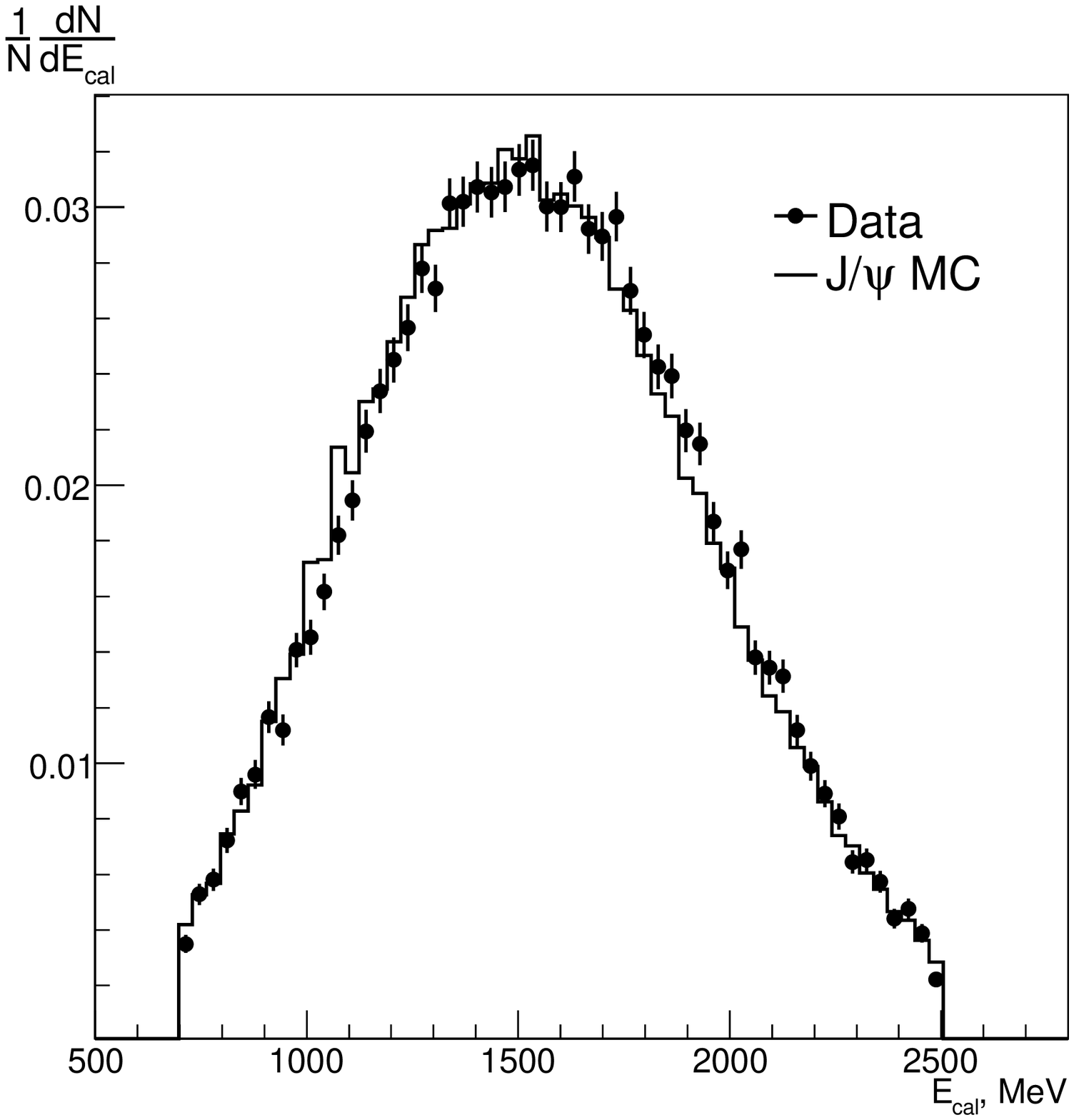}
\hfill
\includegraphics*[width=0.49\textwidth,height=0.25\textheight]{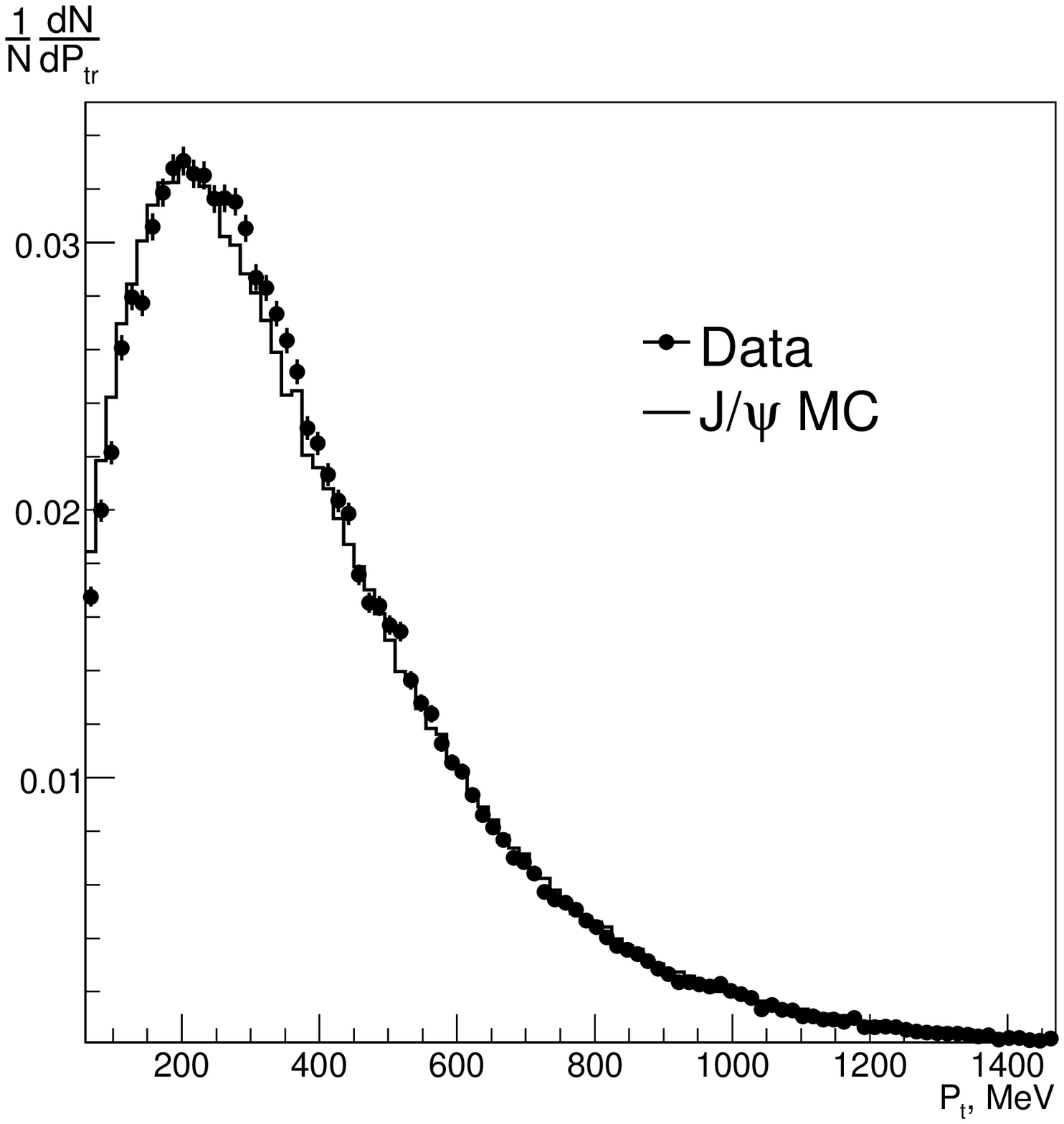}\\
\includegraphics*[width=0.49\textwidth,height=0.25\textheight]{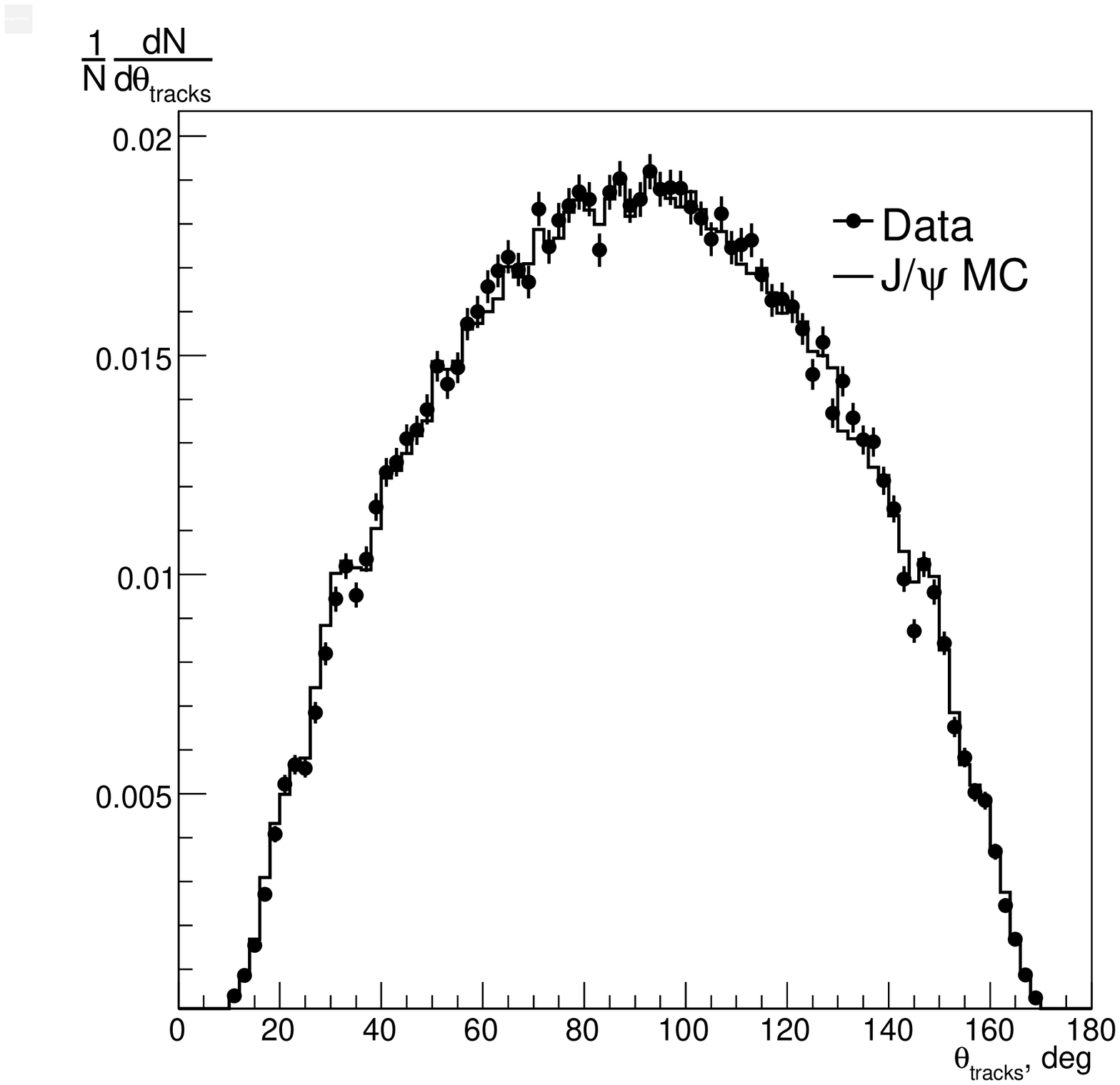}
\hfill
\includegraphics*[width=0.49\textwidth,height=0.25\textheight]{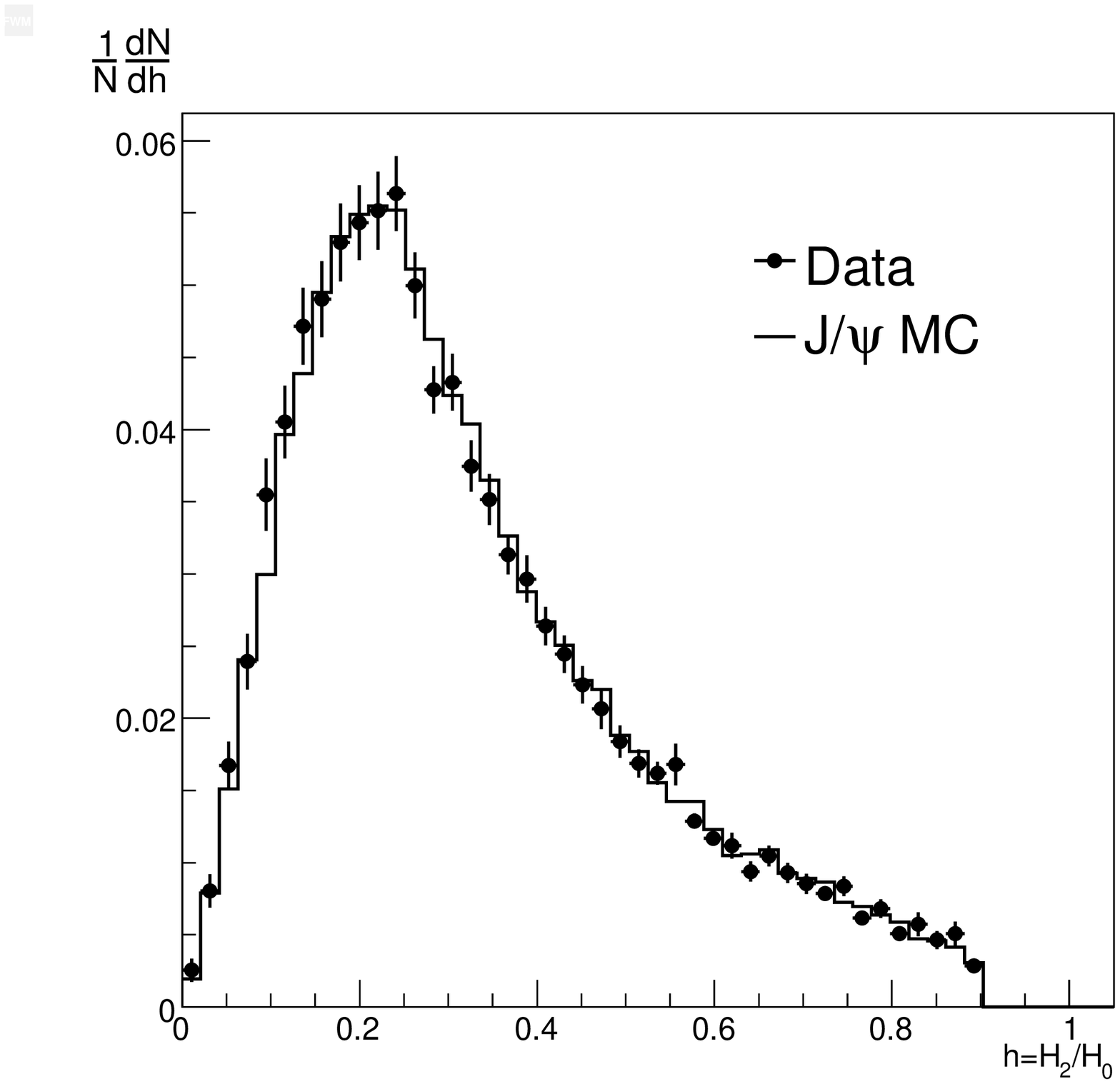}

\caption{Properties of hadronic events produced in the vicinity of the
  $J/\psi$ resonance: the number of tracks from the interaction point
  $N_{\rm IP}$,
  the total number of particles $N_{\rm part}$, energy deposited in the
  calorimeter $E_{\rm cal}$, inclusive $P_{\rm t}$ and $\theta_{\rm tracks}$
  distributions and the ratio of Fox-Wolfram moments $H_2/H_0$. The points
  represent experimental data, the histograms correspond to simulation of $J/\psi$
  decays. All distributions are normalized to unity.}
\label{hadrsel}
\end{figure*}

Figure~\ref{hadrsel} shows comparison between
the most important event characteristics obtained in the experiment
and in the simulation. 

\subsection{Fitting of the data}
\label{subsec:fitting}

We performed a combined fit of the data on
hadron and $e^+e^-$ production  in
the energy range of the $J/\psi$ resonance.

Experimental runs were grouped into points according to run
energy. The collision energy at each point was determined by 
interpolating the beam energy measurements and assuming 
the $e^+e^-$ beam energy symmetry $W=2E_{\text{beam}}$.
A sample of $e^+e^-$ events 
was subdivided into 10 equal angular intervals in the range from 40$^{\circ}$ to 140$^{\circ}$.

The numbers of hadronic $N_i$ and leptonic $n_{ij}$ events observed at
each energy point and each angular interval were
fitted simultaneously  as a function of collision energy and electron
scattering angle using a minimizing function
\begin{equation}
\chi^2 =
\sum_{i}\frac{(N_{i}^{\text{exp}}-N_{i}^{\text{theor}})^2}{N_{i}^{\text{exp}}}+\sum_{i}\sum_{j}
\frac{(n_{ij}^{\text{exp}}-n_{ij}^{\text{theor}})^2}{n_{ij}^{\text{exp}}},
\label{chi2fit}
\end{equation}
where $N_i^{\text{exp/theor}}$ and $n_{ij}^{\text{exp/theor}}$ are experimentally measured and theoretically calculated numbers of hadronic and Bhabha
events, respectively. Theoretically calculated event numbers were
obtained as follows:

\begin{equation}
\begin{split} 
N_{i}^{\text{theor}} = L_i \cdot \sigma^{\text{hadr}}(W_i), \\
n_{ij}^{\text{theor}} = L_i \cdot \sigma^{ee}(W_i, \theta_j).
\end{split} 
\end{equation}
Observed cross sections $\sigma^{\text{hadr}}(W_i)$ and
$\sigma^{ee}(W_i, \theta_j)$ are determined from Eq.~\eqref{BWrelativistic} and Eq.~\eqref{eetoee}, respectively:

\begin{equation}
\begin{split} 
\sigma^{\text{hadr}}(W) = \varepsilon_{\text{hadr}} \int
\sigma^{\text{hadr}}_{\text{n.r.}}(W') G(W, W') dW' + \sigma_{\text{cont}}(W),
\label{shadr}
\end{split} 
\end{equation}
\begin{equation}
\sigma^{ee}(W, \theta) =  \varepsilon_{ee}(\theta)
\left(\frac{d\sigma}{d\theta}\right)^{ee\to  ee}(W) ,
\label{see}
\end{equation}
where the cross section of the annihilation process near the $J/\psi$ 
resonance is convolved with the Gaussian distribution with the energy 
spread $\sigma_W$:

\begin{equation}
 G(W, W') = \frac{g(W-W')}{\sqrt{2\pi}\sigma_W} e^{\textstyle -\frac{(W-W')^2}{2\sigma_W^2}}.
\label{gausw}
\end{equation}
The pre-exponential factor $g$ differs from
unity due to some accelerator-related effects. Its impact on the results
of the measurements is considered in Sec. 5.4. The continuum
cross section is almost constant  
in the vicinity of a
narrow resonance and can be  parametrised with

\begin{equation}
\sigma_{\text{cont}}(W) = \sigma_0 \cdot \left ( \frac{m_{J/\psi}}{W} \right )^2.
\end{equation}

In Eqs.~\eqref{shadr} and \eqref{see}, $\varepsilon_{\text{hadr}}$ and
$\varepsilon_{ee} (\theta)$ are detection efficiencies and their
dependence on beam energy can be neglected. 

Luminosity $L_i$ at $i$-th
energy point was determined as:
\begin{equation}
L_i = R_L \cdot L(E_i),
\end{equation}
where $L(E_i)$ is the integrated luminosity measured by the 
 bremsstrahlung
 luminosity monitor at the $i$-th energy point and  $R_L$ is an
 absolute luminosity calibration factor.

The statistical  uncertainties of parameters
$\Gamma_{ee}(J/\psi)$,
$\Gamma_{ee}(J/\psi) \cdot \mathcal{B}_{ee}(J/\psi)$,
$\Gamma_{ee}(J/\psi) \cdot\mathcal{B}_\text{hadrons}(J/\psi)$
are strongly correlated. To determine these uncertainties accurately,
the fit was performed with two sets of free parameters.
In the first set the parameters 
$\Gamma_{ee}(J/\psi)$ and
$\Gamma_{ee}(J/\psi) \cdot \mathcal{B}_{ee}(J/\psi)$ were floating.
In the second set the parameters
$\Gamma_{ee}(J/\psi) \cdot \mathcal{B}_\text{hadrons}(J/\psi)$ and $\Gamma_{ee}(J/\psi)
\cdot\mathcal{B}_{ee}(J/\psi)$ were floating.
Both sets contained auxiliary free parameters: absolute luminosity calibration
  factor $R_L$, resonance mass $m(J/\psi)$, beam energy
spread $\sigma_W$ and continuum contribution $\sigma_0$. 
To relate the values of $\Gamma_{ee}(J/\psi)$,
$\Gamma_{ee}(J/\psi) \cdot \mathcal{B}_{ee}(J/\psi)$
and $\Gamma_{ee}(J/\psi) \cdot\mathcal{B}_\text{hadrons}(J/\psi)$
the ratio $\Gamma_{e^+e^- }/\Gamma_{\mu^{+}\mu^{-}}(J/\psi) = 1.0022 \pm
0.0065$ was fixed from the KEDR result~\cite{KEDR:2014} and the variation of the ratio
  inside its uncertainties introduces negligible systematic
  uncertainty to the measured values.  The results obtained
  from the fits are listed in Table~\ref{fitrez}. The $J/\psi$  mass value
  is in good  agreement with that published earlier by the KEDR collaboration~\cite{KEDR:2015}.

\begin{table}[hbt!]
\centering
\begin{tabular}{l|l|l}
Free parameter & Fit 1 &  Fit 2 
\\[1mm]\hline
$\Gamma_{ee}\,(\text{keV})$ & $5.550 \pm 0.056$ & -- \\
$\Gamma_{ee} \cdot \mathcal{B}_\text{hadrons}\,(\text{keV})$ &
-- & $4.884 \pm 0.048 $ \\
$\Gamma_{ee} \cdot \mathcal{B}_{ee}\,(\text{keV})$ &  $0.3331 \pm
    0.0066 $ & $0.3331 \pm 0.0066 $ \\
$m\,(\text{MeV})$ & $3096.902 \pm 0.004$ & $3096.902 \pm 0.004$ \\
$R_L$ & $0.973 \pm 0.008$ & $0.973 \pm 0.008$ \\
$\sigma_W \,(\text{MeV})$ & $0.692 \pm 0.004$ & $0.692 \pm 0.004$ \\
$\sigma_0 \,(\text{nb}) $ & $28.70 \pm 1.48$ & $28.70 \pm 1.48$ \\
\hline
\end{tabular}
\caption{Results of two different data fits  performed for $\Gamma_{ee}(J/\psi)$ and $\Gamma_{ee}(J/\psi) \cdot
 \mathcal{B}_\text{hadrons}(J/\psi)$ determination.}
\label{fitrez}
\end{table}

\section{Study of systematic uncertainties}
\label{sec:errors}

Main contributions of systematic uncertainties to the
$\Gamma_{ee}(J/\psi)$, $\Gamma_{ee}(J/\psi)\cdot
 \mathcal{B}_\text{hadrons}(J/\psi)$ and $\Gamma_{ee}(J/\psi) \cdot
  \mathcal{B}_{ee}(J/\psi)$ values  discussed in detail in this section
were merged into five categories: absolute luminosity
measurement, hadron decay simulation, detector and accelerator
effects, theoretical uncertainties.

\subsection{Luminosity uncertainties}
\label{sec:lumerr}

Systematic effects related to luminosity were evaluated by using
variation of the selection requirements.
The requirement on the polar angle  $\theta$ was varied in the broad range, and the
corresponding change in the number of selected Bhabha events reached 50\%. 
All variations are summarized in Table~\ref{eecutsvar} and their
contribution to the total error does not exceed a 0.8\% uncertainty.

\begin{table}[hbt!]
\centering
\begin{tabular}{l|l|c}
Variable & Range variation & Uncertainty,\% 
\\[1mm]\hline
$\theta$  &  $ \theta > 40 \div 50^{\circ}$ and $\theta < 90 \div 140^{\circ}$ & 0.5 \\
$(E_1+E_2)$ & $> 1800 \div 2000 $ MeV & 0.3 \\
$N_{\rm IP}$ definition & $ \rho<0.5 \div 1.0~\text{cm}$, & 0.3 \\
 &  $|\text{z}_0| < 13 \div 55 $ cm  and  & \\
 & $P_t > 0 \div 100 $ MeV &  \\
$E_{\rm cal}-(E_1+E_2)$ & $< 0.1 \div 1.0 \,  E_{\rm cal}$ & 0.3 \\
$\Delta \theta$ acollinearity & $<15 \div 30^{\circ}$  & 0.2 \\
$\Delta \phi$ acollinearity & $< 15 \div 30^{\circ}$ & 0.2 \\
$S_{\rm ch} $ & $< 0.05 \div 0.1 $ & 0.2 \\
\hline
\multicolumn{2}{l|}{Total } & 0.8 \\

\hline
\end{tabular}
\caption{Uncertainties in \% due to variation
 of the selection criteria for $e^+e^-$ events.}
\label{eecutsvar}
\end{table}

In addition, we studied more carefully the following effects. The LKr 
calorimeter was aligned 
to the drift chamber using DC-reconstructed tracks from cosmic events. The position of
the interaction point and the beam-line direction in the
coordinate system of the detector were found using 
the primary-vertex distribution of hadronic events. 

The luminosity uncertainty due to inaccuracy of the alignment was
evaluated  by applying the one-sigma shift during the reconstruction. 
The obtained uncertainty is less than 0.2\%.
The uncertainty due to the imperfect simulation of the calorimeter
response was estimated by varying sensitivity to the energy loss
fluctuations between LKr calorimeter electrodes and appears to be less
than 0.3\%.

The detection efficiency function for electrons,
 $\varepsilon_{ee}(\theta)$, was calculated with 
$J/\psi \to e^+e^-$ simulation, with the $\theta$ angle
measured in the drift chamber or LKr calorimeter, the result difference 
does not exceed 0.3\%. A MC statistical uncertainty corresponds to~0.15\%.
To estimate the uncertainty of the $e^+e^- \to e^+e^-$ scattering cross section,
calculated from Eq.~\eqref{eetoee} two event generators -
BHWIDE and MCGPJ were used. The difference
in the $\Gamma_{ee}(J/\psi) $ value was 0.37\%.

The luminosity spread was estimated as a difference of the results from 
two independent luminosity monitors and was about 0.4\%.
This effect was studied with toy MC 
and the corresponding $\Gamma_{ee}(J/\psi)$ and $\Gamma_{ee}(J/\psi) \cdot
  \mathcal{B}_\text{hadrons}(J/\psi)$ width uncertainties  were  about 0.04\% and $\Gamma_{ee}(J/\psi) \cdot
  \mathcal{B}_{ee}(J/\psi)$ uncertainty  was about 0.06\%.

Sources of the absolute luminosity determination uncertainties are summarized in Table~\ref{lumsyst}.

\begin{table}[htb!]
\centering
\begin{tabular}{l|c}
Source & Uncertainty, \% \\[1mm]\hline
Criterion variation & 0.8 \\
Calorimeter alignment  & 0.2 \\
Calorimeter response & 0.3 \\
Detection efficiency $\varepsilon_{ee}(\theta)$ & 0.3 \\
MC statistics & 0.2 \\
Cross section  & 0.4\\
Relative luminosity  & 0.1 \\
\hline
Total & 1.0 \\
\hline
\end{tabular}
\caption{Systematic uncertainties of the luminosity determination in $\%$. }
\label{lumsyst}
\end{table}

\subsection{Uncertainty due to imperfect simulation of $J/\psi$ decays}
\label{subsec:simulation}

The next important source of uncertainties on the $\Gamma_{ee}(J/\psi)$ value
is the imperfect simulation of $J/\psi$ decays.
To tune the simulation procedure and obtain a reliable
estimate of the systematic uncertainty, we follow the method used in
Ref.~\cite{KEDR:2012}. 

\begin{figure*}[htb!] 
\centering\includegraphics*[width=0.8\textwidth]{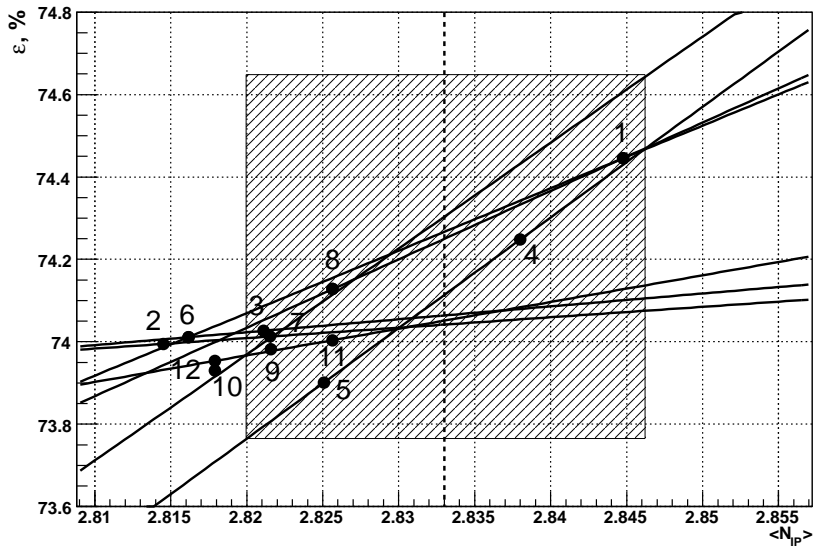}
\caption{Detection efficiency dependence on charged multiplicity for
  different versions of the $J/\psi$ decay simulation. The solid lines
  correspond to variation of the selected parameters. The dotted
  line corresponds to the experimental measured charged multiplicity
  and the rectangular shadow box  shows its statistical error. The vertical limits
  of the shadow box correspond to fit lines crossing the data
  statistical bounds.}
\label{effvsntr}
\end{figure*}

Let us discuss the idea of the method in brief. 
Assume that we have a perfect simulation procedure
capable of reproducing all event characteristics and correlations
between them, but it has a set of internal parameters to be tuned.
By varying one of the parameters, one should trace the change
of the mean value of some observable, for example
the mean multiplicity $\left<N_{\text{\text{IP}}}\right>$, 
and  the detection efficiency $\varepsilon$.
 The simulated value of the observable coincides
with the measured one at the optimal setting of the parameter.
For small variations the detection efficiency linearly depends
 on the mean multiplicity,
therefore the accuracy of the efficiency determination
$\delta\varepsilon =
 \partial\varepsilon/\partial\left<N_{\text{\text{IP}}}\right>\, 
\delta\left<N_{\text{\text{IP}}}\right>$, 
where $\delta\left<N_{\text{\text{IP}}}\right>$ is the 
uncertainty of the experimental value of the multiplicity.
In case of several simulation parameters to vary, one should get the set of
$\varepsilon(N_{\text{\text{IP}}})$ trajectories crossing
together at the point which corresponds to an experimental observable. 
In practice,
the simulation procedure is not perfect, thus
instead of one intersection point
we have the situation depicted in Fig.~\ref{effvsntr}.
The uncertainty
of the detection efficiency grows due to difference in trajectory slopes
obtained with variations of simulation parameters. The estimate of the
uncertainty interval corresponds to the vertical size of the shadow box
in Fig.~\ref{effvsntr} while the horizontal size is determined by the track
multiplicity uncertainty in the experiment.

To obtain the results presented in Fig.~\ref{effvsntr}, 
we iterated as follows: vary one of the 
JETSET parameters and then modify some complementary
parameter to achieve good agreement in observed charged multiplicity.
The values of the mean multiplicity 
and the detection efficiencies obtained for various
settings of parameters are summarized in Table~\ref{gentun}.

The main JETSET parameters to vary are PARJ(21), PARJ(33), 
PARJ(37), PARJ(41) and  PAR(42) referred to
$\sigma_{\rm P_{\text{T}}}$,
$W_{\text{stop}}$, $\delta W_{\text{stop}}$ 
and two parameters $a$ and $b$ for the Lund fragmentation
function, respectively. 
The parameter $\sigma_{\rm P_{\text{T}}}$ is responsible for
a width in the Gaussian transverse-momentum distributions of primary
particles appearing during fragmentation, while  $W_{\text{stop}}$ is the
energy of the jet system below which a final hadron pair is produced. 
This energy is smeared with a relative width $\delta
W_{\text{stop}}$. 
Beside variations of fragmentation function parameters, we tried the fragmentation
with parton showers switched off.

The charged multiplicity was selected for tuning as the most sensitive event
characteristic.
In addition to it, simulated distributions of charged tracks sphericity,
Fox-Wolfram moments,  energy deposited in the calorimeter, the inclusive event
characteristics such as momentum, azimuthal and polar angles,  were
checked for agreement with experimental data. Histogram
shapes were compared using a  Kolmogorov test and simulated samples
that gave the values of the Kolmogorov test lower than 0.6 were rejected.

The multihadron efficiency was averaged over efficiencies corresponding to
an experimentally measured charged multiplicity 
$\left<N_{\text{IP}}\right>$ in Fig.~\ref{effvsntr} 
and equals  $74.2\pm0.4$\%.

\begin{table*}[htb!]
\centering
\begin{tabular}{|l|l|l|l|l|l|l|l|}
\hline
Version & \multicolumn{2}{c|}{JETSET modifications} & $<N_{IP}>$&
$\varepsilon$, \% & $k-test$ & $\chi^2_{\rm MC-data}/ndf$
\\[1mm]\hline
\hline
\multicolumn{7}{|c|}{$\sigma_{\rm P_{\rm T}}$ and  $\delta W_{\rm stop}$ varied} \\
\hline
\  & $\sigma_{P_{T}}$, GeV & $\delta W_{\rm stop}$ &  \multicolumn{4}{c|}{  } \\
\hline
1  & 0.55 & 0.2 & 2.845  & 74.445  &  0.950 & 5.924/7 \\
2  & 0.7 & 0.2   &   2.816 & 74.010  & 0.806 & 9.596/7\\
3 & 0.7 & 0.17 & 2.821  & 74.027  & 0.933 & 7.282/7\\
\hline
\hline
\multicolumn{7}{|c|}{Switching parton showers} \\
\hline
4 & 0.6 & 0.2  & 2.838 & 74.248 & 0.996 & 4.387/7 \\
5$^*$ & 0.6 & 0.2 &   2.825  & 73.901 & 0.999 & 6.697/7  \\
\hline
\hline
\multicolumn{7}{|c|}{$W_{\rm stop}$ varied} \\
\hline
\  & $\sigma_{P_{T}}$, GeV & $W_{\rm stop}$, GeV & \multicolumn{4}{c|}{  }\\
\hline
6 & 0.65 & 0.56    & 2.815  & 73.995  &  0.663 & 8.497/7 \\
7 & 0.65 & 0.52    & 2.822 & 74.013  & 0.903 & 4.761/7 \\
\hline
\hline
\multicolumn{7}{|c|}{Fragmentation function with $a$=0.2, $b$=0.58} \\
\hline
\  & $\sigma_{\rm P_{\rm T}}$, GeV & $\delta W_{\rm stop}$, GeV &
\multicolumn{4}{c|}{  } \\
\hline
8 & 0.65 & 0.2 & 2.826  & 74.128 & 0.954 & 8.574/7 \\
9  & 0.65 & 0.17 &  2.822  & 73.982 &  0.839 & 13.288/7 \\
10 & 0.7 &  0.2 & 2.818 & 73.930  & 0.685 & 11.234/7 \\
\hline
\hline
\multicolumn{7}{|c|}{Parameters of fragmentation function varied} \\
\hline
\  & $a$ & $b$ & \multicolumn{4}{c|}{  } \\
\hline
11  & 1.0 & 0.7  & 2.826 &  74.004 & 0.979 & 10.483/7 \\
12  & 0.5 & 0.65 &  2.818 & 73.954  &  0.986  & 9.514/7 \\
\hline
\multicolumn{7}{l}{\footnotesize{$^*$ Switched-off parton shower}}
\end{tabular}
\caption{Comparison of different versions of MC simulation for $J/\psi$ decays. 
  JETSET modification \  parameters are presented. For each
  simulated sample, the detection efficiency was calculated  and results of
  Kolmogorov and $\chi^2$ tests on the charged multiplicity distribution
  are shown as well as average value. }
\label{gentun}
\end{table*}

For the calculation of the mean multiplicity some track selection
criteria are required. Their choice leads  
to an additional uncertainty on the detection efficiency
which is smaller than 0.3\%.
The track reconstruction efficiency  is not exactly
the same for the experimental data and simulation.
The difference was studied using Bhabha events and cosmic
tracks  and the appropriate correction was introduced in the detector
simulation with an uncertainty smaller than 0.1\%.

For reweighting we used significant and well-measured $J/\psi$ decay
branching fractions. To check a systematic uncertainty, the remaining
branching fractions were added to the list and corresponding MC event 
weights were recalculated. This leads to uncertainty of less than $0.1\%$ 
on the measured $\Gamma_{ee}(J/\psi)$ and $\Gamma_{ee}(J/\psi) \cdot
  \mathcal{B}_\text{hadrons}(J/\psi)$ values.  

All systematic uncertainties due to imperfect simulation are
summarized in Table~\ref{simsyst}. 

\begin{table}[htb!]
\centering
\begin{tabular}{l|c}
Source & Uncertainty, \% \\[1mm]\hline
Generator  & 0.6 \\
Track selection & 0.3 \\
MC statistics & 0.3 \\
Tracking efficiency  & 0.1 \\
PDG  branchings & 0.1 \\
\hline
Total & 0.7 \\
\hline
\end{tabular}
\caption{Systematic uncertainties of $\Gamma_{ee}(J/\psi) \cdot
  \mathcal{B}_\text{hadrons}(J/\psi)$ due to imperfect simulation of $J/\psi$ decays.}
\label{simsyst}
\end{table}

\subsection{Detector-related uncertainties}
\label{subsec:detector}

The major sources of the detector-related systematic uncertainties in
the $\Gamma_{ee}(J/\psi)$ width are listed in
Table~\ref{detsyst}. 

\begin{table}[htb!]
\centering
\begin{tabular}{l|c}
Source & Uncertainty, \% \\[1mm]\hline
Criterion variation & 0.5 \\
Cosmic suppression & 0.3 \\
Nuclear interaction & 0.2 \\
Tracking $P_{\rm t}/\theta$ resolution & 0.2 \\
Trigger efficiency  & 0.5 \\
\hline
Total & 0.8 \\
\hline
\end{tabular}
\caption{Sources of detector-related systematic uncertainties in \%.}
\label{detsyst}
\end{table}

The effects of  possible sources of the detector-related
uncertainties were evaluated by varying the event selection requirements. 
 Minimum and maximum total energies
deposited in the calorimeter were varied to 500 and 2700 MeV, respectively. 
A requirement on the Fox-Wolfram moments was removed from
selection. A requirement on the number of tracks from interaction points was
tightened to have $N_{\rm IP} > 1$ and track selections on $\rho$,  $\text{z}_0$ 
and $P_{\rm t}$ were varied and
the obtained difference did not exceed 0.2\%.
The results are presented
in Table~\ref{hadrcutsvar} giving in total about 0.5\%.
\begin{table}[h!]
\centering
\begin{tabular}{l|l|c}
Variable & Range variation & Uncertainty, \% 
\\[1mm]\hline
$E_{\rm cal}$ & $ E_{\rm cal} > 500 \div 700 $ and  & 0.3\\
 & $ E_{\rm cal} < 2500 \div 2700$ MeV & \\
$E_{\rm LKr}$/$E_{\rm cal}$ & $>0 \div 0.15 $& 0.3\\
$N_{\rm IP}$ & $ \rho<0.5 \div 1.5~\text{cm}$,  $\text{z}_0<13 \div 18~\text{cm}$,  & 0.2 \\
      & $P_{\rm t}>60 \div 100~\text{MeV}$, $N_{\rm IP} > 1$  &   \\
$H_2/H_0$& $<0.9 \div 1$ & 0.1 \\
\hline
\multicolumn{2}{l|}{Total} & 0.5 \\
\hline
\end{tabular}
\caption{$\Gamma_{ee}(J/\psi)\cdot
  \mathcal{B}_\text{hadrons}(J/\psi)$ uncertainty in \% due to variation
  of the selection criteria for hadronic events.}
\label{hadrcutsvar}
\end{table}

To estimate the uncertainty related to the cosmic background, 
the condition on the muon system veto was replaced with a condition 
on the average ToF time with the number of hits in the muon system 
not larger than two. The difference was found to be about 0.3\% for
$\Gamma_{ee}(J/\psi)$ and  $\Gamma_{ee}(J/\psi) \cdot \mathcal{B}_{\rm hadrons}(J/\psi)$
and 0.1\%
for $\Gamma_{ee}(J/\psi) \cdot \mathcal{B}_{ee}(J/\psi)$.

In addition, we used two models of nuclear interaction
during simulation - with the
GHEISHA~\cite{gheisha} and FLUKA~\cite{fluka} packages as they are implemented
in GEANT 3.21. 
The variation of the resulting $\Gamma_{ee}$ value was
about 0.2\%.

The two methods to achieve data and MC agreement in the momentum
 resolution and the angular resolution were used: we scale either the
 assumed systematic uncertainties of x(t) determination or the
 spatial resolution of the drift chamber. That gives a 0.2\% systematic
 uncertainty.

The trigger inefficiency includes three contributions.
The inefficiency of time-of-flight counters used in the primary trigger
was determined with especially selected $e^+e^- \to e^+e^-$ and cosmic
events and equals $0.3\%$.
A systematic uncertainty 
due to crosstalk in VD electronics was evaluated
as a difference of results with two sets of VD simulation parameters
obtained by using cosmic and Bhabha events. It was about 0.2\%.
And the last effect is a veto from CsI
crystals near the beam line, which was estimated by varying
corresponding trigger thresholds and equals 0.3\%.

\subsection{Accelerator uncertainties}
\label{subsec:collider}

The influence of the machine background was estimated by using a data set
collected with separated beams. The number of hadronic events selected
from this data set was rescaled to the full data sample proportionally to the
integrals of the beam currents. The contribution of background events to
the observed cross section is about 6-12 nb.
The number of selected hadron events was corrected for 
the number of estimated background events and the data were refitted.
The relative
uncertainty does not exceed 0.2\%.

The non-Gaussian effects in the total collision energy distribution
contribute about 0.2\% to the $\Gamma_{ee}(J/\psi)$ uncertainty. To
estimate this contribution, we added a pre-exponential factor 
while
convolving the cross section with a Gaussian function in Eq.~\eqref{gausw} (details are discussed in~\cite{KEDR:2015}):
\begin{equation}
g(\Delta W)=(1+a \cdot \Delta W+b \cdot \Delta W^2)/(1+b \cdot \sigma_W^2).
\end{equation}

To check the uncertainty related to the beam energy determination,
the values of energy assigned to the data points were corrected
within their errors using the known shape of the resonance cross section.
For that, eleven free parameters $E^{\rm fit}_i$ were introduced in the
fit function~\eqref{chi2fit} and the compensating term,
\begin{equation}
\chi^2_E = \sum_i
\frac{\left(E^{\rm fit}_i-E_i\right)^2}{\sigma_{\text{E}_i}^2},
\end{equation}
was added, 
where $E_i$ is the energy obtained from interpolation
of the resonance depolarization data and
$\sigma_{{\rm E}_i}$ is its estimated accuracy.
For the cross section calculations, the values $W_i=2E^{\rm fit}_{i}$
were used.
The variation of the fitted $\Gamma_{ee}$ value was about 0.3\%.

The list of accelerator uncertainties is presented in Table~\ref{accelsyst}.

\begin{table}[htb!]
\centering
\begin{tabular}{l|c}
Source & Uncertainty, \%  \\[1mm]\hline
Collider background & 0.2 \\
Non-Gaussian energy & 0.2 \\
Beam energy  & 0.3 \\
\hline
Total & 0.4 \\
\hline
\end{tabular}
\caption{Accelerator-related uncertainties contributions in \%.}
\label{accelsyst}
\end{table}

\subsection{Other uncertainties}
\label{subsec:othererr}

The interference parameter $\lambda$ in the fit was fixed at the
value of 0.39 assuming that the sum in \eqref{eq:lambdaSum} vanishes.
To verify the uncertainty related to this parameter, we left $\lambda$
floating resulting in a shift of 0.2\% on the
$\Gamma_{ee}(J/\psi) $ and $\Gamma_{ee}(J/\psi) \cdot \mathcal{B}_\text{hadrons}(J/\psi)$ values
and about 0.1\% on the $\Gamma_{ee}(J/\psi) \cdot
\mathcal{B}_{ee}(J/\psi)$ value.

Deviation of $\tilde{\Gamma}_{\rm h}$ from a sum of partial hadronic
widths $\Gamma_{\rm hadrons}$ due to interference effects was estimated in the
Bayesian approach under the assumption that all phases in
Eq.~\eqref{eq:lambdaSum} have equal probability as discussed in~\cite{KEDR:2012}. At
the fitted value $\lambda=0.36 \pm 0.14$, the effect does not exceed 0.3\%.

The accuracy of the analytic expression~\eqref{BWrelativistic} is
about 0.1\%. In addition, the 0.1\% accuracy of the radiative-correction calculation~\cite{KF} should be taken into account. The inaccuracy of
simulation of FSR effects with PHOTOS is negligible in our analysis.

The sum in
quadrature of all contributions listed in this subsection is about 0.4\%.

\section{Summary}
\label{sec:summary}

The parameters of the $J/\psi$ meson have been measured by using the data
collected with the KEDR detector at the VEPP-4M $e^+ e^-$
collider. Two data fits were performed, the first one was used to
obtain partial lepton widths $\Gamma_{ee}(J/\psi) \cdot
\mathcal{B}_\text{hadrons}(J/\psi)$ and $\Gamma_{ee}(J/\psi) \cdot
\mathcal{B}_{ee}(J/\psi)$. Their errors are strongly correlated,
therefore to obtain the total leptonic width $\Gamma_{ee}(J/\psi)$ a separate
fit was introduced.  Our results are
$$
 \Gamma_{ee}(J/\psi) = 5.550 \pm 0.056 \pm 0.089 \, \text{keV},
$$
$$
\Gamma_{ee}(J/\psi) \cdot \mathcal{B}_\text{hadrons}(J/\psi) = 4.884 \pm
    0.048 \pm 0.078 \, \text{keV}, 
$$
$$
\Gamma_{ee}(J/\psi) \cdot \mathcal{B}_{ee}(J/\psi) = 0.3331 \pm
    0.0066 \pm 0.0040 \, \text{keV}.
$$

The first and second uncertainties are statistical and systematic,
respectively. The major sources of the systematic uncertainties for the
$\Gamma_{ee}(J/\psi)$ and  $\Gamma_{ee}(J/\psi) \cdot
\mathcal{B}_\text{hadrons}(J/\psi)$ values are summarized in
Table~\ref{allsyst} and  the total systematic uncertainty equals 1.6\%. 
For the  $\Gamma_{ee}(J/\psi) \cdot \mathcal{B}_{ee}(J/\psi)$ product, 
the total systematic uncertainty equals 1.2\%.

\begin{table}[htb!]
\centering
\begin{tabular}{l|c|c|c}
Source & \multicolumn{3}{c}{Uncertainty, \%} \\\cline{2-4} 
 & $\Gamma_{ee}$  & $\Gamma_{ee} \cdot
\mathcal{B}_\text{hadrons}$ & $\Gamma_{ee}\cdot
\mathcal{B}_{ee}$ 
\\[1mm]\hline
Luminosity  & 1.0 & 1.0 & 1.0 \\
Simulation of $J/\psi$ decays & 0.7 & 0.7 & -- \\
Detector response & 0.8 & 0.8 & 0.4 \\
Accelerator-related effects & 0.4 & 0.4 & 0.4\\
Theoretical uncertainties & 0.4 & 0.4 & 0.2 \\
\hline
Total & 1.6 & 1.6 & 1.2 \\
\hline
\end{tabular}
\caption{Dominant systematic uncertainties in the
  $\Gamma_{ee}(J/\psi)$, $\Gamma_{ee}(J/\psi) \cdot  \mathcal{B}_\text{hadrons}(J/\psi)$ and  $\Gamma_{ee}(J/\psi) \cdot  \mathcal{B}_{ee}(J/\psi)$  values.}
\label{allsyst}
\end{table} 

\begin{figure*}[htb!] 
\centering\includegraphics*[width=0.45\textwidth,height=0.37\textheight]{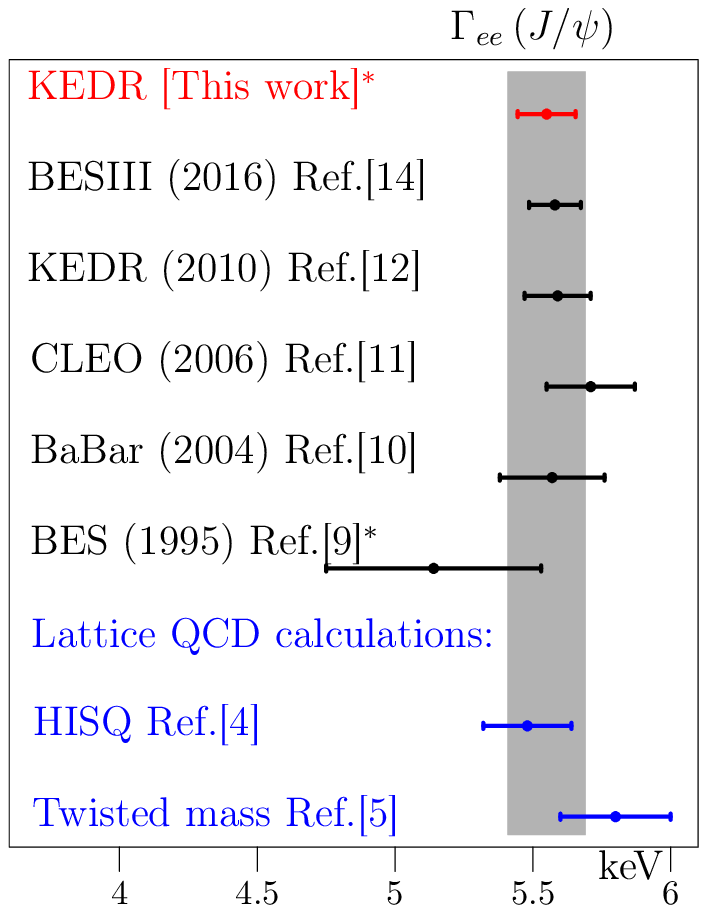}
\hfill
\centering\includegraphics*[width=0.45\textwidth,height=0.37\textheight]{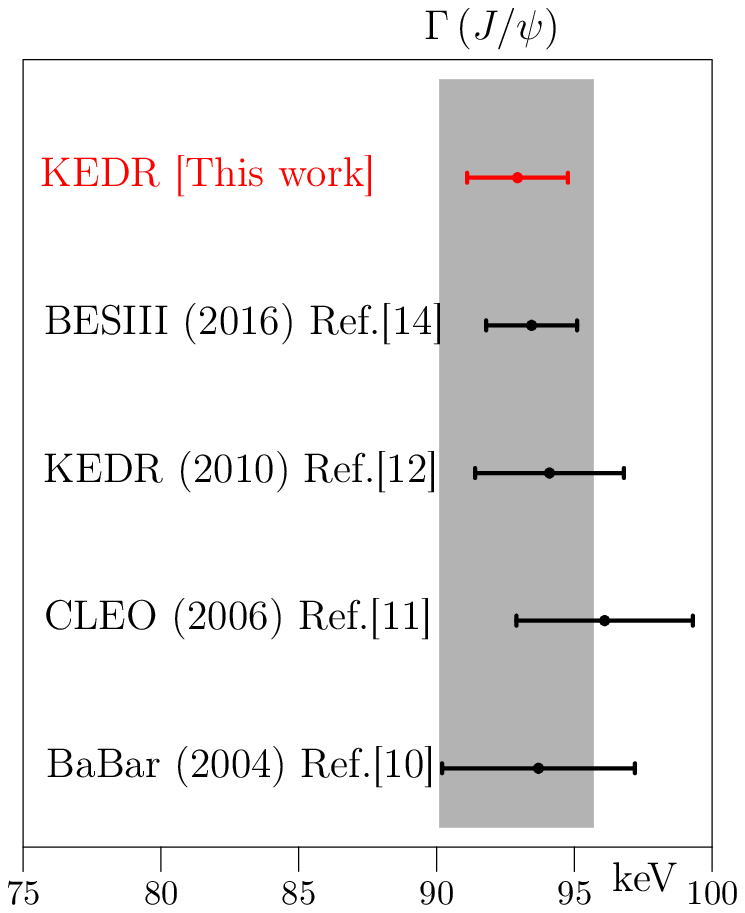}\\
\raggedright\footnotesize{$^*$ Direct measurement}
\caption{Comparison of $\Gamma_{ee} (J/\psi)$ and $\Gamma (J/\psi)$
  measured in the most precise experiments and $\Gamma_{ee} (J/\psi)$
  predictions from lattice QCD calculations. The $\Gamma (J/\psi)$ value from
  the BESIII experiment was calculated from~\cite{bes16} using the world-average lepton branching
  fraction~\cite{pdg2014}. The gray band corresponds to the
world-average value with allowance for the uncertainty in it.}
\label{gee}

\end{figure*}

Our result for the $\Gamma_{ee}(J/\psi) \cdot
  \mathcal{B}_\text{hadrons}(J/\psi)$ value  is consistent with and four
  times more precise than the previous direct measurement in the hadronic channel~\cite{bes95}. The obtained $\Gamma_{ee}(J/\psi) \cdot
  \mathcal{B}_{ee}(J/\psi)$ value is in good agreement with our
  previous measurement~\cite{KEDR:2010} and supersedes it. 

Taking into account
$\mathcal{B}_{ee}(J/\psi) = (5.971 \pm 0.032)\%$
from~\cite{pdg2014} we determined the total width of the $J/\psi$ meson:
$$
\Gamma  = 92.94 \pm 1.83 \, \text{keV}.
$$

The leptonic and total widths of the $J/\psi$ meson $\Gamma_{ee}$ are
known from the 
BESIII\cite{bes16}, CLEO~\cite{cleo06} and BaBar~\cite{babar04}
experiments.
The values were calculated from $\Gamma_{ee}(J/\psi) \cdot
\mathcal{B}_{\mu\mu}(J/\psi)$ measured in  the radiation process
 $e^+ e^- \rightarrow \mu^+ \mu^- \gamma$ with the 
$J/\psi$ meson decaying to muon pair.  

The electronic and total widths  obtained in our analysis  agree well with 
the world average   
$\Gamma_{ee}= 5.55\pm 0.14 \pm 0.02$~keV and $\Gamma = 92.9 \pm
2.8$~keV \cite{pdg2014}.
Figure~\ref{gee} represents comparison of our $\Gamma_{ee}(J/\psi)$ and
$\Gamma(J/\psi)$ results with those obtained in previous
experiments.

\section*{Acknowledgments}
We greatly appreciate the efforts of the staff of VEPP-4M to provide
good operation of the accelerator complex.

This work was supported by Russian Science Foundation under
project N 14-50-00080.  Work related to $J/\psi$ Monte Carlo generator tuning was
partially supported by Russian Foundation for Basic Research under
grant 12-02-01076-a.

\end{document}